\documentclass[a4paper]{article}

\usepackage{INTERSPEECH2022}

\usepackage{xcolor}
\usepackage{comment}

\def\cameraReadyChanges{\textcolor{black}}

\title{Dyadic Interaction Assessment from Free-living Audio for Depression Severity Assessment}
\name{Bishal Lamichhane$^1$,Nidal Moukaddam$^2$, Ankit B. Patel$^1$, Ashutosh Sabharwal$^1$}
\address{
  $^1$ECE, Rice University, Houston, Texas, USA\\
  $^2$Menninger Department of Psychiatry, Baylor College
of Medicine, Houston, Texas, USA.}
\email{\{bishal.lamichhane,ankit.patel,ashutosh.sabharwal\}@rice.edu, Nidal.Moukaddam@bcm.edu}

\begin{document}

\maketitle
\begin{abstract}

Psychomotor retardation in depression has been associated with speech timing changes from dyadic clinical interviews. In this work, we investigate speech timing features from free-living dyadic interactions. Apart from the possibility of continuous monitoring to complement clinical visits, a study in free-living conditions would also allow inferring sociability features such as dyadic interaction frequency implicated in depression. We adapted a speaker count estimator as a dyadic interaction detector with a specificity of 89.5\% and a sensitivity of 86.1\% in the DIHARD dataset. Using the detector, we obtained speech timing features from the detected dyadic interactions in multi-day audio recordings of 32 participants comprised of 13 healthy individuals, 11 individuals with depression, and 8 individuals with psychotic disorders. The dyadic interaction frequency increased with depression severity in participants with no or mild depression, indicating a potential diagnostic marker of depression onset. However, the dyadic interaction frequency decreased with increasing depression severity for participants with moderate or severe depression. In terms of speech timing features, the response time had a significant positive correlation with depression severity. Our work shows the potential of dyadic interaction analysis from audio recordings of free-living to obtain markers of depression severity.
  
\end{abstract}


\section{Introduction}

Major depressive disorder (MDD), or depression, is one of the most common mental health disorders~\cite{WHO2021Depression} and remains underdiagnosed~\cite{argyropoulos2015depressive,basta2021frequency}. 
%Though depression remains largely treatable~\cite{kroenke2001phq}, it is still underdiagnosed~\cite{argyropoulos2015depressive,basta2021frequency}.
Currently, questionnaires such as PHQ~\cite{spitzer1999validation} and CES-D~\cite{radloff1977ces} are used as a diagnostic tool for depression. These questionnaires rely on subjective symptom reporting from the individuals and could lead to biased or incorrect assessments. Several recent works have thus investigated objective markers of depression based on mobility~\cite{saeb2015}, communication logs~\cite{cao2020tracking}, brain imaging~\cite{yoshida2017prediction}, social media posts~\cite{de2013predicting}, etc. 
 
Speech is one of the commonly investigated modalities to identify objective markers of depression. Psychomotor retardation in depression reportedly alters acoustic and timing features (turn-taking behaviors) in one's speech which has been explored as potential markers of depression severity~\cite{cummins2015review}. For example, the authors in~\cite{afshan2018effectiveness} found voice quality features to improve depression severity prediction. Similarly, the authors in~\cite{yamamoto2020} analyzed dyadic clinical interviews and found the pause time (time between utterances of the interviewee) and the response time (time between interviewer and interviewee's utterance) was longer for the depressed participants compared to healthy participants.

Currently, speech timing features as depression severity markers are obtained from clinical interviews~\cite{yamamoto2020,cannizzaro2004voice}. If such features could be extracted from free-living conversations, frequent monitoring would be possible to complement the assessments from the clinical visits. As the speech timing-based depression severity markers have been obtained from dyadic interactions (interviews) in existing studies, we aim to identify dyadic interactions in the free-living audio and study speech timing features in such interactions. 

Identifying dyadic interactions in free-living audio could be challenging due to background noise, intra-speaker speech variabilities, the spontaneity of conversations, etc. No existing studies have investigated the dyadic interaction detection task. But several studies have addressed audio-based speaker counting~\cite{xu2013crowd++,wang2020speaker} which is closely related. A dyadic interaction detection, after all, is a binarized version of speaker counting (if the number of speakers is two or not). Speaker counting pipelines' ability to detect dyadic interactions, however, is also yet to be quantified. Similarly, though the findings of altered speech iteration and varying prosody are already documented from controlled studies, these have not been studied well in free-living with participants representing a wide depression severity range. Additionally, though isolation as a symptom of worsening depression is understood, it is not known whether it affects all sorts of interactions such as dyadic.

Our work is the first demonstration of the feasibility of dyadic interaction analysis from multi-day free-living audio for mental health applications;  we make the following two main contributions.
First, we develop a dyadic interaction detector based on a speaker counting pipeline and demonstrate the feasibility of dyadic interaction detection in diverse audio recording settings. We adapted ECoNet~\cite{lamichhane2022}, a speech processing pipeline to assess everyday conversational networks from free-living audio, for dyadic interaction detection. ECoNet deals with the challenges encountered in free-living audio by, for example, using a robust voice activity detection model (VAD) and identifying spuriously detected speakers due to background noise and intra-speaker speech variations. ECoNet-based dyadic detector outperformed a baseline model of dyadic interaction detection using VAD outputs, indicating the significance of speaker inference for the detection task. 

Our second contribution is dyadic interaction analysis in multi-day audio recordings of free-living from a diverse participant group, including a clinical population. We used a wrist-worn wearable to obtain continuous audio recordings, for up to a week, from 32 participants which included healthy, depressed, and individuals with psychotic disorders. From the dataset obtained, we analyzed the relation of the dyadic interaction frequency and speech timing features with depression severity. Our results are in line with the dyadic interaction hypothesis reported in previous work~\cite{Elmer2020} for a comparable participant group; the frequency of dyadic interactions increased with increasing depression severity. Thus, we provide an audio-based sensing alternative to RFID-based sensing proposed in~\cite{Elmer2020} which required all individuals in a dyadic pair to have worn an RFID tag. The audio-based sensing only requires the \textit{target individual} to wear the sensor. We additionally found that the association between dyadic interaction frequency and depression severity is reversed for the moderate and severely depressed participant groups which were not represented in the earlier study~\cite{Elmer2020}. In terms of speech timing features, we found that response time has a significant positive correlation with depression severity, similar to the observations from controlled studies.  

\section{Dyadic Interaction Detection}

\subsection{ ECoNet-based Detector}

\begin{figure*}[!t]
    \centering
    \includegraphics[width=1.6\columnwidth]{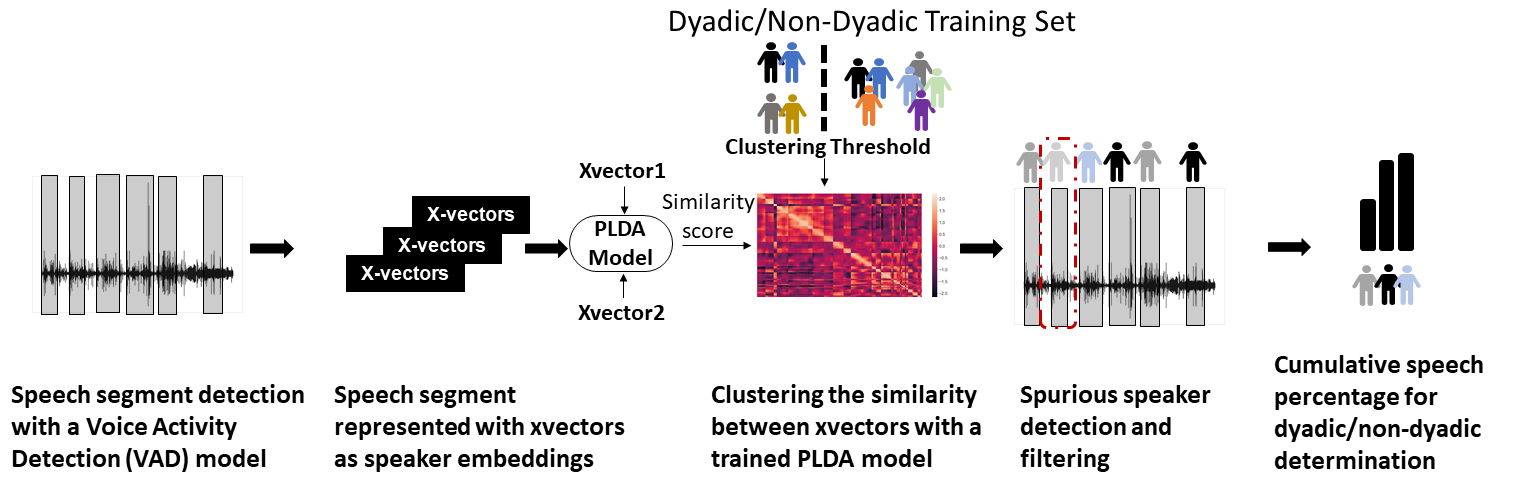}
    \caption{ECoNet pipeline proposed in~\cite{lamichhane2022} to estimate conversational network from free-living audio was adapted for dyadic interaction detection.}
    \label{fig:econet_pipeline}
\end{figure*}

We developed a dyadic interaction detector based on the everyday conversational network estimator (ECoNet) proposed in~\cite{lamichhane2022}. ECoNet was adapted for the dyadic interaction detection by binarizing the inferred conversational network size as 2 (dyadic) or not 2 (non-dyadic). The architecture of ECoNet is given in Figure~\ref{fig:econet_pipeline}. It consists of a SincNet-LSTM-based VAD~\cite{Bredin2019} to identify speech segments, xvectors~\cite{Snyder2018} as speaker embeddings, clustering for unsupervised speaker identification/count, and a random forest-based machine learning model for spurious speaker detection. \cameraReadyChanges{The ECoNet pipeline has the clustering threshold as a hyperparameter which was tuned for the dyadic interaction detection task in the DIHARD development set (Section~\ref{sec:dihard_evaluation}).}

\subsection{Baseline Model Using VAD Features}

A dyadic interaction could also be possibly detected from the properties of the identified speech segments with VAD. An audio recording with a higher number of speakers could have higher variability in speech segment duration, characterizing the inter-speaker variability of typical utterance lengths. Similarly, there might be differences in inter-segment duration due to the different number of constituent speakers. Accordingly, we trained a Random Forest (RF)-based dyadic interaction detector using the following four features: mean and standard deviation of speech segment length; mean and standard deviation of inter-segment duration. \cameraReadyChanges{The number of estimators in RF was set to 51; a higher number of estimators did not result in improved cross-validation performance in the DIHARD development set (Section~\ref{sec:dihard_evaluation}) in an increment of 10 estimators starting with 11 estimators.} 

\subsection{Evaluation of Dyadic Interaction Detectors}
\label{sec:dihard_evaluation}

We trained and evaluated the dyadic interaction detectors on the DIHARD development (DIHARD-DEV) and evaluation (DIHARD-EVAL) dataset~\cite{ryant2021dihard} respectively. DIHARD dataset consists of audio sequences from diverse settings such as restaurant conversations, courtroom recordings, conversations in clinics, meetings, etc. Each audio sequence is nominally 10 minutes long, though there are some shorter sequences also, and has speaker diarization annotations, i.e., the speech segments have corresponding speaker labels. 

\begin{comment}
We show example annotations from the DIHARD dataset sequences with different numbers of speakers in Figure~\ref{fig:dihard_diffspeakers}. Speakers in a sequence could have different speech percentages and be active throughout the sequence or in certain segments only.   

\begin{figure}[!htb]
    \centering
    \includegraphics[width=\columnwidth]{fig/different_speaker_interactions_segment.png}
    \caption{Examples of different interactions, with each speaker represented by a color, in the segments of sequences in the DIHARD dataset. The dataset consists of audio recordings from different scenarios and thus varies in the number of speakers present across sequences. The top plot shows a segment with multiple speakers. Speaker1 is present throughout the segment and is the dominant speaker in turn-taking. The middle plot shows an interaction with annotations for 3 speakers. However, Speaker3 has only a small number of speech segments assigned to it and could likely be incorrectly annotated. The bottom plot shows a dyadic interaction with two speakers taking turns in interactions.}
    %\textcolor{red}{this figure/visualization does not provide any value. I don't know what to look for}}
    
    %Bishal -> Added this figure to give some idea to readers about different kinds of N-speaker interactions in the DIHARD dataset. I added some clarification in the caption to describe the interaction but I guess we can skip this plot altogether.
    \label{fig:dihard_diffspeakers}
\end{figure}
\end{comment}

We define a sequence in the DIHARD dataset as dyadic if most interactions ($>$90\%) are attributed to exactly two speakers. This allows to include some sequences that have been incorrectly annotated with new speakers assigned to a few speech segments of existing speakers or have annotations with insignificant speakers (e.g., background, non-involved speakers) as dyadic. Further, we only include sequences longer than 5 minutes since we aim to extract speech timing features from longer conversations only, as employed in previous studies in controlled settings~\cite{yamamoto2020}.   

We trained ECoNet by identifying the clustering threshold resulting in the highest dyadic interaction detection accuracy in the DIHARD-DEV dataset. Then, the trained model is evaluated in the independent DIHARD-EVAL dataset. The results obtained is shown in Table~\ref{tab:Dyadic_detector_dihard}. A high accuracy of 86.1\% for dyadic interaction detection in the DIHARD-EVAL dataset is obtained. The spurious speaker detection module helped improve the detection accuracy. \cameraReadyChanges{Naturally, the accuracy is higher in the DIHARD-DEV dataset where the clustering threshold was tuned.} 
%The specificity of the trained dyadic interaction detector on the DIHARD-EVAL set is high, as required in our application to identify dyadic interactions from where speech timing features will be extracted. 
%Still, some sequences which were non-dyadic were detected as dyadic. Most of these sequences were found to be noisy or have background speakers in a monologue sequence included within the DIHARD dataset. 
The baseline RF-based model was also trained on the DIHARD-DEV dataset and evaluated in the DIHARD-EVAL dataset. The accuracy obtained with the baseline model was lower compared to the ECoNet-based detector, indicating the importance of speaker inference for dyadic interaction detection.

\begin{table}[!htb]
    \centering
    \caption{Evaluation of ECoNet-based and RF-based baseline model for dyadic interaction detection in the DIHARD dataset}
    \begin{tabular}{|c|c|c|c|}
        \hline
         \textit{Dataset} & \textbf{Accuracy} & \textbf{Specificity} & \textbf{Sensitivity}\\
            \hline
           \multicolumn{4}{|c|}{With Spurious Speaker Detection} \\
           \hline
           DIHARD-DEV  & \textbf{98.2\%} & 94.9\% & 99.4\%  \\
           \hline
           DIHARD-EVAL & \textbf{86.1\%}  & 89.5\% & 83.4\%  \\
           \hline
           \multicolumn{4}{|c|}{Without Spurious Speaker Detection} \\
           \hline
           DIHARD-DEV  & 83.5\% & 79.6\% & 84.9\%  \\
           \hline
           DIHARD-EVAL & 83.4\%  & 71.9\% & 87.3\%  \\
           \hline
           \multicolumn{4}{|c|}{Baseline RF Model} \\
           \hline
           DIHARD-EVAL  & 80.3\% & 73.1\% & 82.4\%  \\
           \hline
           
    \end{tabular}
\label{tab:Dyadic_detector_dihard}
\end{table}

\section{Free-living Audio Dataset Analysis}

We used the ECoNet-based dyadic interaction detector to analyze the multi-day free-living audio from a diverse participant group. Particularly, we inferred where dyadic interaction happens and obtained speech timing features in the detected dyadic interactions using the speaker diarization output from ECoNet. Then, we assessed the relation of dyadic interaction frequency and speech timing features with depression severity.  

\subsection{Data Collection}

\begin{figure}[!htb]
  \centering
  \includegraphics[width=\linewidth]{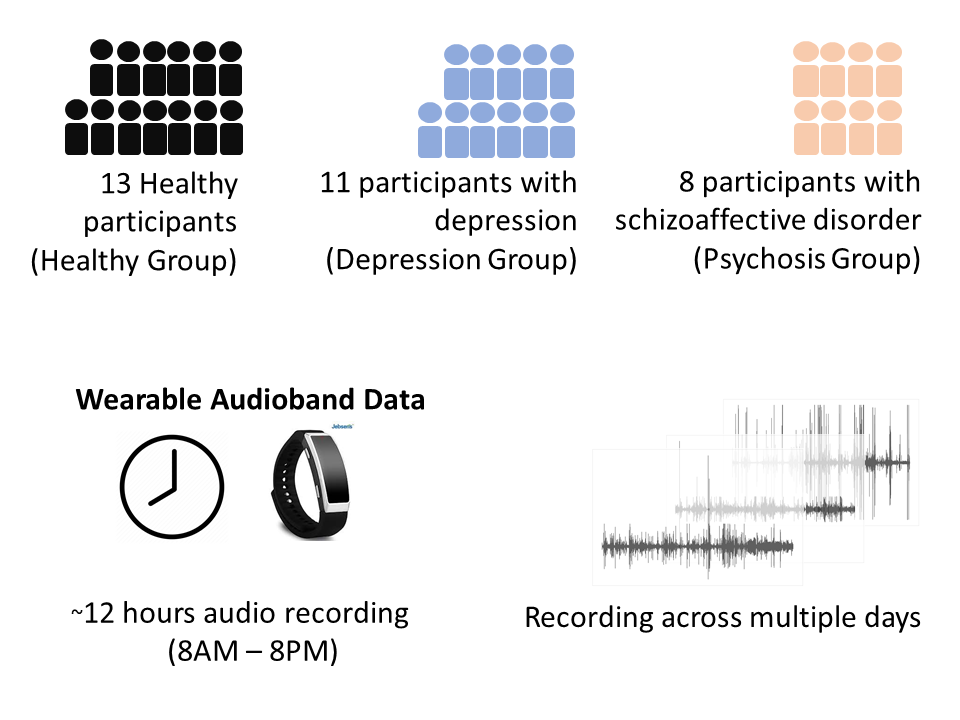}
  \caption{Continuous audio recordings of free-living, from up to a week, were obtained from 32 participants with diverse mental health conditions using a wrist-worn audio recorder.}
  \label{fig:study_data}
\end{figure}

We conducted a study with 32 participants comprising 13 healthy individuals, 11 individuals with major depressive disorders, and 8 individuals with psychotic disorders, as depicted in Figure~\ref{fig:study_data}. For each participant, we obtained continuous audio recordings during the day, for up to 7 days, using a wrist-worn audio recorder. The audio dataset thus obtained is referred to as the free-living audio dataset. We assessed the depression severity of the participant using the 9-item Patient Health Questionnaire (PHQ-9)~\cite{spitzer1999validation}. The study was approved by the Institutional Review Board (IRB) at Rice University, Harris Health System, and Baylor College of Medicine.  

\subsection{Dyadic Ratio}

\begin{figure}
    \centering
    \includegraphics[width=\columnwidth]{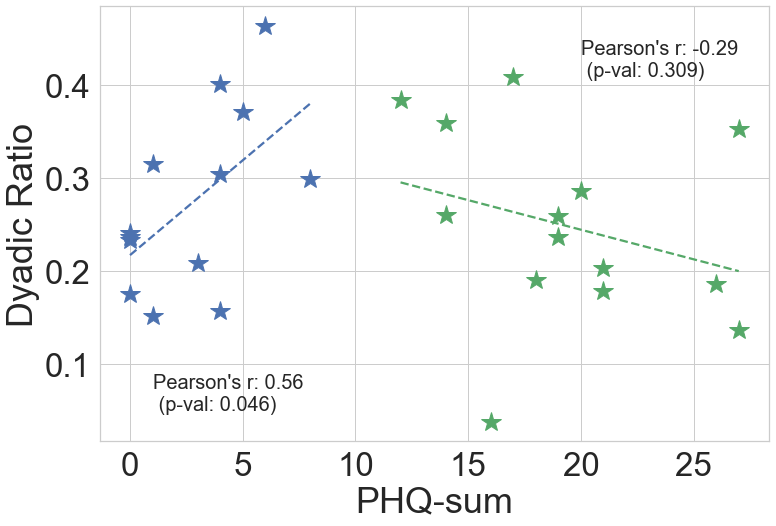}
    \caption{Dyadic ratio and PHQ-9 scores for participants in the free-living audio dataset}
    \label{fig:dyadic_ratio_phq}
\end{figure}

We evaluated each 10-minute non-overlapping sliding window of the audio in the free-living audio dataset as being a dyadic interaction or not. The number of dyadic interaction windows compared to the total number of windows for a participant is defined as the dyadic ratio, representing the dyadic interaction frequency. The relation of the dyadic ratio with the participants' PHQ-9 scores is shown in Figure~\ref{fig:dyadic_ratio_phq}. The dyadic ratio increased with increasing PHQ-9 scores (Pearson's correlation coefficient: 0.56, p-value: 0.046) but reduced as the PHQ-9 scores are higher (correlation coefficient: -0.29, p-value:0.31).

%A similar pattern was obtained when the normalization factor was the number of windows with $>2$ speakers (likely group interactions); the correlation coefficient was 0.42 (p-value: 0.15) for PHQ-9 $<$ 10 and -0.39 (p-value: 0.17) for PHQ-9 scores $\geq$ 10.  

\subsection{Speech Timing Features}

%Segments from example dyadic interaction windows detected from the free-living audio dataset are shown in Figure~\ref{fig:clinical_dyadic_example}. Some interactions are speech-rich (containing a high percentage of speech segments), while others could have less speech content overall, indicating less active interactions. To extract speech timing features, it would be desirable to consider speech-rich interactions which would match the scenarios in a clinical interview setting. Thus, we extracted speech timing features from the top-10 windows with the highest speech percentages from each participant's dyadic interaction windows, averaging the features from different windows. However, the \textit{target speaker}, i.e., the participant, among the two speakers in each dyadic interaction, has to be identified first.  

%camera -ready changes --> deleting figure4 altogether
\cameraReadyChanges{Among the dyadic interaction windows detected, some interactions are speech-rich (containing a high percentage of speech segments) while others could have less speech content overall, indicating less active interactions.} To extract speech timing features, it would be desirable to consider speech-rich interactions which would match the scenarios in a clinical interview setting. Thus, we extracted speech timing features from the top-10 windows with the highest speech percentages from each participant's dyadic interaction windows, averaging the features from different windows. However, the \textit{target speaker}, i.e., the participant, among the two speakers in each dyadic interaction, has to be identified first.  

\begin{comment}
\begin{figure}
    \centering
    \includegraphics[width=0.8\columnwidth]{fig/baylor_dyadicseq_2min.png}
    \caption{Segments from example dyadic sequences detected in the free-living audio dataset. Sequences could have different percentages of speech depending upon the interaction type.} %\textcolor{red}{Not  sure what the colors mean  - need legends in the figure}}
    %Bishal -> Added labels for speaker
    \label{fig:clinical_dyadic_example}
\end{figure}
\end{comment}

\begin{figure}
    \centering
    \includegraphics[width=\columnwidth]{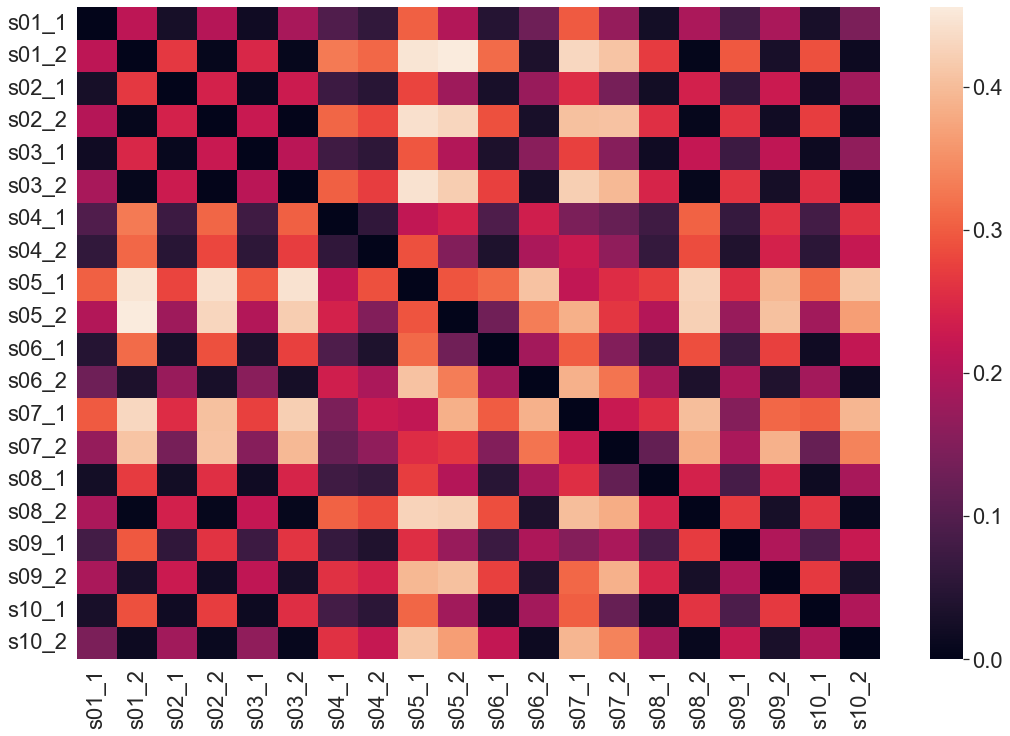}
    \caption{ \cameraReadyChanges{Heatmap showing the distances between speakers across dyadic interactions for an example participant in the free-living audio dataset. sXX\_1, sXX\_2 (e.g., s01\_1, s01\_2) indicates the two speakers in a given dyadic interaction sXX (e.g., s01). A checkerboard-like pattern is seen in the distance heatmap, indicating that a common speaker is likely present in all/most of the dyadic interactions.}}
    \label{fig:clinical_heatmap}
\end{figure}

For all speakers in the dyadic interactions from a participant's audio recordings, we computed their average speaker embeddings from corresponding speech segments. Then, we obtained the cosine distances between the embeddings. Figure~\ref{fig:clinical_heatmap} shows an example of such distances from a participant's audio recordings. \cameraReadyChanges{A checkerboard-like pattern is seen in the inter-speaker distance map. This pattern would arise when a common speaker is present across interactions. The common speaker would have a small distance (high similarity) between their embeddings, appearing as a dark square in the example heatmap. We assume the common speaker to be the \textit{target speaker}, i.e., the participant.}

%, and identify them as the speaker present in the speaker combinations across dyadic interactions with the lowest sum of distances among them.}

%If $s[X] \in \{s[01],s[02],...,s[10]\}$ are the dyadic interactions identified for a participant, and $s[X]_1$ and $s[X]_2$ represent the two speakers in a dyadic interaction, a list of target speakers from each dyadic interaction $T = {T_1,T_2,...,T_10}$, $T_i \in {1,2}$ is obtained as min T j 1 to 10, k 1 to 2 dist(s[j]_,s[j]_)

\begin{figure}[!htb]
    \centering
    \includegraphics[width=\columnwidth]{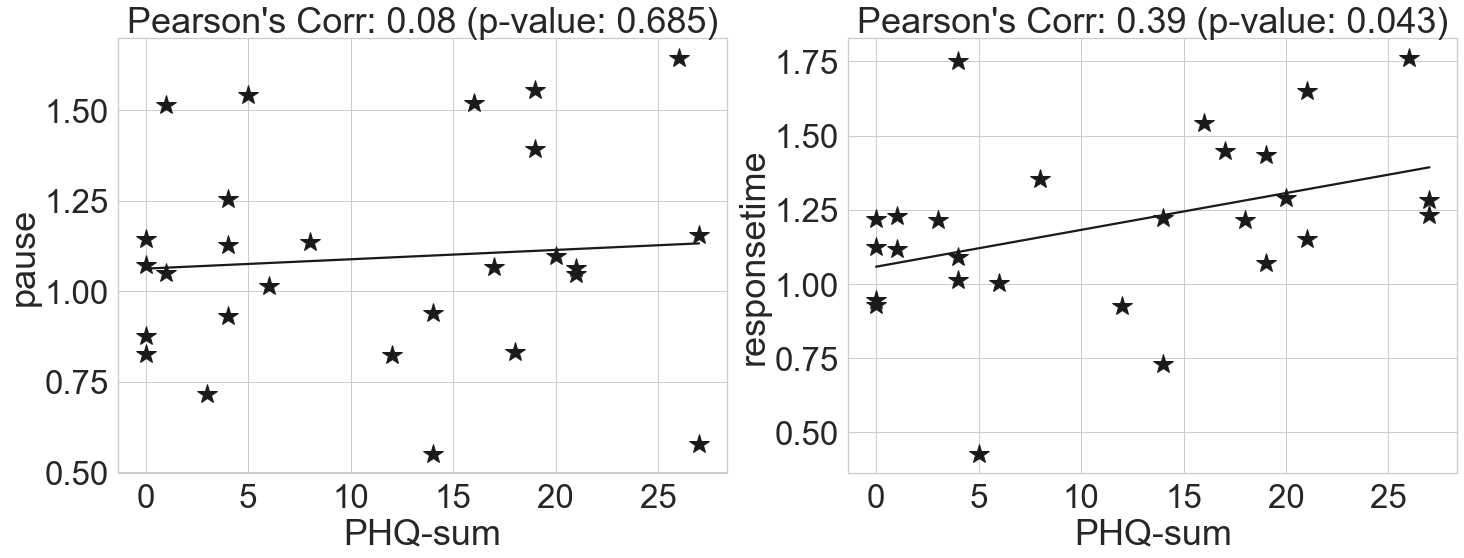}
    \caption{Speech timing features obtained from dyadic interactions detected in the free-living audio dataset compared to the PHQ-9 scores of the participants.}
    \label{fig:speechtiming_phq}
\end{figure}

For the identified target speaker, we assessed the pause time, i.e., the average time between the speaker's utterances and response time, i.e., the average time between the other speaker's and the target speaker's utterances. The average speech timing features compared to the participant's PHQ-9 scores are shown in Figure~\ref{fig:speechtiming_phq}. The response time correlated significantly with the PHQ-9 score (Pearson's correlation coefficient of 0.39). Of the 9 items in PHQ-9, the group difference of response time between participants with a score of 0 for a particular item and those with a score $>$ 0 was highest for the $1^{st}$ item \textit{Little interest or pleasure in doing things?} (t-stats: 2.17, p-value: 0.039), followed by the $9^{th}$ item \textit{Thoughts that you would be better off dead, or thoughts of hurting yourself in some way?} (t-stats: 2.10 , p-value: 0.045), and the $2^{nd}$ item \textit{Feeling down, depressed, or hopeless?} (t-stats: 1.89 , p-value: 0.068). Higher response time could hence be driven by the general lack of interest in activities when in a depressed state. The group difference for the $8^{th}$ item \textit{Moving or speaking so slowly that other people could have noticed? Or so fidgety or restless that you have been moving a lot more than usual?}, \cameraReadyChanges{chosen for analysis because the item directly implicates speech behaviors}, was not significant (t-stats: 1.36, p-value: 0.184). The group with non-zero scores on the $8^{th}$ item, though, had a higher average response time (1.28 seconds) than participants with a score of 0 (1.11 seconds). 

%We also analyzed the differences in PHQ-9 scores and average response time between the three groups represented in the free-living audio dataset as shown in Figure~\ref{fig:groupdiff_responsetime}. The depression and psychosis group had significantly higher PHQ-9 scores than the healthy group. The psychosis group had lower average PHQ-9 scores than the depression group but had a higher response time, though the differences were not significant.

\cameraReadyChanges{We also analyzed the differences in PHQ-9 scores and average response time between the three groups represented in the free-living audio dataset. The depression and psychosis group had significantly higher PHQ-9 scores (18.1$\pm$6.5 and 13.9$\pm$9.9 respectively) than the healthy group (2.9$\pm$2.6). The psychosis group had a higher response time (1.29$\pm$0.23 seconds) than the depression group (1.21$\pm$0.28) but the difference was not significant.}

%camera-ready version changes, removing this figure
\begin{comment}
\begin{figure}[!htb]
    \centering
    \includegraphics[width=\columnwidth]{fig/groupdiff_phq_responsetime.png}
    \caption{Group difference in terms of their PHQ-9 score (left) and average response time in a dyadic interaction (right)}
    \label{fig:groupdiff_responsetime}
\end{figure}
\end{comment}

\section{Discussion}

Identification and characterization of dyadic interactions from free-living audio could provide depression severity markers. Dyadic interaction frequency represents a sociability behavior. Similarly, the speech timing features from dyadic interactions could model the psychomotor retardation associated with depression. We evaluated a dyadic interaction detector based on a speaker counting architecture in this work. The detector had a high accuracy in the DIHARD dataset and was used to identify instances of dyadic interaction in a free-living audio dataset of multi-day audio recordings from 32 participants with diverse mental health conditions. We identified the target speaker in each dyadic interaction and obtained their speech timing features.

We obtained a positive association between dyadic ratio and depression severity in participants with no or mild depression, i.e., for participants with PHQ-9 scores $<$ 10 (Figure~\ref{fig:dyadic_ratio_phq}). This result is in line with the dyadic interaction hypothesis proposed in~\cite{Elmer2020}. Individuals with higher depression severity tend to have more dyadic interactions, e.g., to ruminate about their worries/concerns. The number of unique dyadic partners for a depressed individual could still be less, reflecting a lower total social network size, and needs to be investigated in future work. The study of~\cite{Elmer2020} consisted of healthy participants with only $\approx$15-20\% of them with some depression; participants with higher depression severity were not represented. We found that the association between the dyadic ratio and depression severity is rather negative for the moderately or severely depressed participants (PHQ-9 scores $\geq$ 10). This association could reflect the tendency of highly depressed individuals to avoid any form of interaction. Our results, though preliminary, indicate that the dyadic ratio could be useful only as a \cameraReadyChanges{pre-diagnostic}/monitoring feature for healthy individuals prone to depression. \cameraReadyChanges{Further, non-linear functions would be required to model depression severity from features such as dyadic ratio.}

In terms of the speech timing features, we found that the response time obtained from the dyadic interactions in free-living is positively correlated with depression severity (Figure~\ref{fig:speechtiming_phq}). This observation is in line with previous studies that have reported increasing response time with higher depression severity~\cite{yamamoto2020,nilsonne1988speech, alpert2001reflections, yang2012detecting}. Though pause time has also been reported to correlate positively with depression severity in earlier studies~\cite{yamamoto2020}, we did not obtain similar observations in our analysis. The dyadic interactions of free-living could be eliciting different speech timings compared to clinical interviews, leading to this discrepancy, and need further investigation.

\cameraReadyChanges{All three participant groups had good adherence to the
study. The average per-participant audio data duration was 49.16, 53.27, and 53.75 hours in the healthy, depression, and psychosis groups, respectively. The psychosis group had a slightly
lower average number of recording days (4.87 days) than
the healthy and depression group (5.53 days, and 6.09 days
respectively) but the difference was not significant.}

%%Removed this based on Reviewer 2 feedback.. difference in number of days between psychosis and healthy group had p-val 0.055 (just on borderline) but not significant.

%this could be due to the illness of the psychosis
%group or technological difficulties (e.g., forgetting to charge the
%recorder overnight). Future studies could assess the acceptance,
%comfort, and adherence to speech technologies in groups with
%different mental health conditions.

Our study shows how audio from free-living could be leveraged to obtain dyadic interaction-based markers of depression severity. These markers could complement inferences obtained from controlled conditions such as clinical interviews and allow more continuous monitoring. However, audio recording in free-living also raises questions about wider acceptability due to privacy concerns. Our analysis does not parse any spoken content or attempt to identify speakers' true identities which could help retain some privacy. Future work should further investigate the trade-offs between privacy/acceptability and benefits in health monitoring inherent in speech technologies. Continuous audio monitoring-based health applications could gain wider acceptance by improving the technology to preserve user privacy by design and providing compelling (health) applications.

%All figures must be centered on the column (or page, if the figure spans both columns). Figure captions should follow each figure and have the format given in Figure~\ref{fig:speech_production}.
%Figures should be preferably line drawings. If they contain gray levels or colors, they should be checked to print well on a high-quality non-color laser printer.

%$Graphics (i.\,e., illustrations, figures) must not use stipple fill patterns because they will not reproduce properly in Adobe PDF. Please use only SOLID FILL COLORS.

%Figures which span 2 columns (i.\,e., occupy full page width) must be placed at the top or bottom of the page.

%\subsection{Hyperlinks}
%For technical reasons, the proceedings editor will strip all active links from the papers during processing. Hyperlinks can be included in your paper, if written in full, e.\,g.\ ``http://www.foo.com/index.html''. The link text must be all black. 
%Please make sure that they present no problems in printing to paper.

%The reference format is the standard IEEE one. References should be numbered in order of appearance, for example \cite{Davis80-COP}, \cite{Rabiner89-ATO}, \cite[pp.\ 417--422]{Hastie09-TEO}, and \cite{YourName21-XXX}.

\section{Acknowledgements}
BL was partially supported by the Ken Kennedy Institute 2021/22 Energy HPC Conference Graduate Fellowship.

\bibliographystyle{IEEEtran}

\bibliography{interspeech}

\begin{thebibliography}{10}
\providecommand{\url}[1]{#1}
\csname url@samestyle\endcsname
\providecommand{\newblock}{\relax}
\providecommand{\bibinfo}[2]{#2}
\providecommand{\BIBentrySTDinterwordspacing}{\spaceskip=0pt\relax}
\providecommand{\BIBentryALTinterwordstretchfactor}{4}
\providecommand{\BIBentryALTinterwordspacing}{\spaceskip=\fontdimen2\font plus
\BIBentryALTinterwordstretchfactor\fontdimen3\font minus
  \fontdimen4\font\relax}
\providecommand{\BIBforeignlanguage}[2]{{%
\expandafter\ifx\csname l@#1\endcsname\relax
\typeout{** WARNING: IEEEtran.bst: No hyphenation pattern has been}%
\typeout{** loaded for the language `#1'. Using the pattern for}%
\typeout{** the default language instead.}%
\else
\language=\csname l@#1\endcsname
\fi
#2}}
\providecommand{\BIBdecl}{\relax}
\BIBdecl

\bibitem{WHO2021Depression}
\BIBentryALTinterwordspacing
``Depression.'' [Online]. Available:
  \url{https://www.who.int/news-room/fact-sheets/detail/depression}
\BIBentrySTDinterwordspacing

\bibitem{argyropoulos2015depressive}
K.~Argyropoulos, C.~Bartsokas, A.~Argyropoulou, P.~Gourzis, and E.~Jelastopulu,
  ``Depressive symptoms in late life in urban and semi-urban areas of
  {South-West Greece}: An undetected disorder?'' \emph{Indian journal of
  psychiatry}, vol.~57, no.~3, p. 295, 2015.

\bibitem{basta2021frequency}
M.~Basta, K.~Micheli, P.~Simos, I.~Zaganas, S.~Panagiotakis, K.~Koutra,
  C.~Krasanaki, C.~Lionis, and A.~Vgontzas, ``Frequency and risk factors
  associated with depression in elderly visiting primary health care {(PHC)}
  settings: {Findings} from the {Cretan} aging cohort,'' \emph{Journal of
  Affective Disorders Reports}, vol.~4, p. 100109, 2021.

\bibitem{spitzer1999validation}
R.~L. Spitzer, K.~Kroenke, and J.~B. Williams, ``Validation and utility of a
  self-report version of {PRIME-MD}: The {PHQ} primary care study.''
  \emph{JAMA: Journal of the American Medical Association}, 1999.

\bibitem{radloff1977ces}
L.~S. Radloff, ``The {CES-D} scale: {A} self-report depression scale for
  research in the general population,'' \emph{Applied psychological
  measurement}, vol.~1, no.~3, pp. 385--401, 1977.

\bibitem{saeb2015}
S.~Saeb, M.~Zhang, C.~J. Karr, S.~M. Schueller, M.~E. Corden, K.~P. Kording,
  and D.~C. Mohr, ``Mobile phone sensor correlates of depressive symptom
  severity in daily-life behavior: an exploratory study,'' \emph{Journal of
  medical Internet research}, vol.~17, no.~7, p. e175, 2015.

\bibitem{cao2020tracking}
J.~Cao, A.~L. Truong, S.~Banu, A.~A. Shah, A.~Sabharwal, N.~Moukaddam
  \emph{et~al.}, ``Tracking and predicting depressive symptoms of adolescents
  using smartphone-based self-reports, parental evaluations, and passive phone
  sensor data: development and usability study,'' \emph{JMIR mental health},
  vol.~7, no.~1, p. e14045, 2020.

\bibitem{yoshida2017prediction}
K.~Yoshida, Y.~Shimizu, J.~Yoshimoto, M.~Takamura, G.~Okada, Y.~Okamoto,
  S.~Yamawaki, and K.~Doya, ``Prediction of clinical depression scores and
  detection of changes in whole-brain using resting-state functional {MRI} data
  with partial least squares regression,'' \emph{PloS one}, vol.~12, no.~7, p.
  e0179638, 2017.

\bibitem{de2013predicting}
M.~De~Choudhury, M.~Gamon, S.~Counts, and E.~Horvitz, ``Predicting depression
  via social media,'' in \emph{Seventh international AAAI conference on weblogs
  and social media}, 2013.

\bibitem{cummins2015review}
N.~Cummins, S.~Scherer, J.~Krajewski, S.~Schnieder, J.~Epps, and T.~F.
  Quatieri, ``A review of depression and suicide risk assessment using speech
  analysis,'' \emph{Speech communication}, vol.~71, pp. 10--49, 2015.

\bibitem{afshan2018effectiveness}
A.~Afshan, J.~Guo, S.~J. Park, V.~Ravi, J.~Flint, and A.~Alwan, ``Effectiveness
  of voice quality features in detecting depression,'' \emph{Interspeech 2018},
  2018.

\bibitem{yamamoto2020}
M.~Yamamoto, A.~Takamiya, K.~Sawada, M.~Yoshimura, M.~Kitazawa, K.-c. Liang,
  T.~Fujita, M.~Mimura, and T.~Kishimoto, ``Using speech recognition technology
  to investigate the association between timing-related speech features and
  depression severity,'' \emph{PloS one}, vol.~15, no.~9, p. e0238726, 2020.

\bibitem{cannizzaro2004voice}
M.~Cannizzaro, B.~Harel, N.~Reilly, P.~Chappell, and P.~J. Snyder, ``Voice
  acoustical measurement of the severity of major depression,'' \emph{Brain and
  cognition}, vol.~56, no.~1, pp. 30--35, 2004.

\bibitem{xu2013crowd++}
C.~Xu, S.~Li, G.~Liu, Y.~Zhang, E.~Miluzzo, Y.-F. Chen, J.~Li, and B.~Firner,
  ``Crowd++ unsupervised speaker count with smartphones,'' in \emph{Proceedings
  of the 2013 ACM international joint conference on Pervasive and ubiquitous
  computing}, 2013, pp. 43--52.

\bibitem{wang2020speaker}
W.~Wang, F.~Seraj, N.~Meratnia, and P.~J. Havinga, ``Speaker counting model
  based on transfer learning from {SincNet} bottleneck layer,'' in \emph{2020
  IEEE International Conference on Pervasive Computing and Communications
  (PerCom)}.\hskip 1em plus 0.5em minus 0.4em\relax IEEE, 2020, pp. 1--8.

\bibitem{lamichhane2022}
B.~Lamichhane, N.~Moukaddam, A.~B. Patel, and A.~Sabharwal, ``Econet:
  Estimating everyday conversational network from free-living audio for mental
  health applications,'' \emph{IEEE Pervasive Computing}, 2022.

\bibitem{Elmer2020}
T.~Elmer and C.~Stadtfeld, ``{Depressive symptoms are associated with social
  isolation in face-to-face interaction networks},'' \emph{Scientific reports},
  vol.~10, no.~1, p. 1444, 2020.

\bibitem{Bredin2019}
H.~Bredin, R.~Yin, J.~M. Coria, G.~Gelly, P.~Korshunov, M.~Lavechin, D.~Fustes,
  H.~Titeux, W.~Bouaziz, and M.-P. Gill, ``Pyannote. audio: neural building
  blocks for speaker diarization,'' in \emph{ICASSP 2020-2020 IEEE
  International Conference on Acoustics, Speech and Signal Processing
  (ICASSP)}.\hskip 1em plus 0.5em minus 0.4em\relax IEEE, 2020, pp. 7124--7128.

\bibitem{Snyder2018}
D.~Snyder, D.~Garcia-Romero, G.~Sell, D.~Povey, and S.~Khudanpur,
  ``{X-Vectors}: {Robust} {DNN} embeddings for speaker recognition,'' in
  \emph{2018 IEEE International Conference on Acoustics, Speech and Signal
  Processing (ICASSP)}, 2018, pp. 5329--5333.

\bibitem{ryant2021dihard}
N.~Ryant, P.~Singh, V.~Krishnamohan, R.~Varma, K.~Church, C.~Cieri, J.~Du,
  S.~Ganapathy, and M.~Liberman, ``The third dihard diarization challenge,''
  \emph{arXiv preprint arXiv:2012.01477}, 2020.

\bibitem{nilsonne1988speech}
A.~Nilsonne, ``Speech characteristics as indicators of depressive illness,''
  \emph{Acta Psychiatrica Scandinavica}, vol.~77, no.~3, pp. 253--263, 1988.

\bibitem{alpert2001reflections}
M.~Alpert, E.~R. Pouget, and R.~R. Silva, ``Reflections of depression in
  acoustic measures of the patient’s speech,'' \emph{Journal of affective
  disorders}, vol.~66, no.~1, pp. 59--69, 2001.

\bibitem{yang2012detecting}
Y.~Yang, C.~Fairbairn, and J.~F. Cohn, ``Detecting depression severity from
  vocal prosody,'' \emph{IEEE transactions on affective computing}, vol.~4,
  no.~2, pp. 142--150, 2012.

\end{thebibliography}

\end{document}